\documentclass[amsmath, amssymb, preprintnumbers, noshowpacs, 
showkeys,aps,superscriptaddress,twocolumn]{revtex4}

\usepackage{mathrsfs}
\usepackage{color} 
\usepackage{hyperref}
\usepackage{slashed}
\usepackage{graphicx}
\usepackage{latexsym}
\usepackage{epstopdf}
\usepackage{enumitem}
\usepackage{multirow}
\usepackage{mathtools}
\usepackage{bbm}
\usepackage{bm}
\usepackage{amsfonts}
\usepackage{hyperref}
\usepackage{braket}
\usepackage{verbatim}
\usepackage{subfigure}

\newcommand{\beq}{\begin{equation}}
\newcommand{\eeq}{\end{equation}}

\newcommand{\be}{\begin{equation}}
\newcommand{\ee}{\end{equation}}

\begin{document}

\title{Effective field theory of bosons with finite-range interaction 
\\ in a disordered environment}

\author{Alberto Cappellaro}
\email{cappellaro@pd.infn.it}
\affiliation{Dipartimento di Fisica e Astronomia ``Galileo Galilei''
and CNISM, Universit\`a di Padova, via Marzolo 8, 35131 Padova, Italy}
\author{Luca Salasnich}
\affiliation{Dipartimento di Fisica e Astronomia ``Galileo Galilei''
and CNISM, Universit\`a di Padova, via Marzolo 8, 35131 Padova, Italy}
\affiliation{CNR-INO, via Nello Carrara, 1 - 50019 Sesto Fiorentino, Italy}
\date{\today{}}

\begin{abstract}

We investigate the low-temperature properties of a ultracold gas 
made of bosonic alkali-metal atoms with finite-range interaction 
under the effect of a disordered environment. 
The statistical characterization of the disorder is 
investigated within an effective-field-theory formalism 
for a generic spatial dimension $d$. Moving to $d=3$, where 
all the arising divergences are properly regularized, 
we focus on the depletion of both the condensate 
and superfluid densities. At zero temperature we obtain 
meaningful analytical formulas for the condensate fraction 
and the superfluid fraction which take into account 
the interplay among scattering length, effective range, and disorder strength.
\end{abstract}


\maketitle

\section{Introduction}

During the last decades, a great interest has been aroused by the interplay 
between interactions and disorder. 
The investigation of quantum effects in many-particle systems
moving in disordered environment goes back in time up to the
prediction of a localized phase for free electrons in random-lattice structures
 \cite{anderson-1958}.

Similarly to other condensed matter issues, significant efforts  
have been devoted to explore the role of a disordered environment 
in the framework of ultracold atomic gases \cite{lewenstein-review,bloch-review}.
The main reason lies in the remarkable experimental control 
over the relevant physical parameters such as 
densities and interaction strengths. 
In order to construct a random potential mimicking a porous medium, 
a viable strategy consists in superimposing two different 
optical lattices \cite{damski-2003,schulte2005}. 
Experimentally, another possibility is provided by laser-speckle fields, 
arising from the interference pattern of waves with the same frequency 
but different phases and amplitudes \cite{dainty-book,lye2005,clement2005}.

In this paper we focus on disordered bosons within their superfluid 
phase but in presence of a finite-range interaction between the atoms. 
By taking into account a non-local two-body potential, 
nonuniversal corrections to the thermodynamic potential 
can be obtained beyond the usual mean-field picture
\cite{braaten-1999,braaten-2001,cappellaro-2017}.
As a consequence, detachments from universality may represent a viable 
approach to reveal the elusive contribution of quantum fluctuations. 
These corrections can also 
be included in a modified Gross-Pitaevskii equation (GPE), 
leading to nonuniversal dynamical features
\cite{gao-2003,zinner2009,veksler2014,sgarlata-2015,heinonen2019}.

Our analysis is carried on within the framework of the 
effective-field-theory, where one can relate the coupling constants 
of the Euclidean action functional to the measurable s-wave scattering 
parameters, via the T-matrix technique \cite{braaten-2001,stoof-book}. 
From a technical point of view, in order to implement a quantum field 
theory in presence of a disorder (external) potential, 
different strategies are at disposal. 
Since we are interested in how disorder 
and finite-range interactions both contribute to modify the condensate 
and superfluid depletion, 
we adopt a perturbative approach: the disorder is assumed to be weak, 
in a similar way as quantum and thermal fluctuations. Their role is then 
taken into account up to the one-loop (i. e. Gaussian) level.
For a zero-range interaction, 
it has been shown that also a very weak disorder 
affects the thermodynamic picture of a bosonic 
superfluid \cite{giorgini1994,falco2007,falco2007-2}.
 
On the other hand, the case of strong disorder has to be treated 
non-perturbatively.
The most standard approaches all rely upon the so-called replica trick, 
first proposed in the context of spin glasses 
\cite{anderson1975,mezard-book,parisi1990,mezard1991}.
In \cite{graham2009}, a replicated Hartree-Fock theory managed to 
reproduce the phase diagram for
interacting bosons where also the localized phase is included. 
However, in this paper we are going to focus only on the weak disorder 
regime, where
the replica trick is not crucial, since it basically reproduces the 
perturbative results,
as shown for dirty superconductors \cite{nelson1990} and cold atoms 
\cite{cappellaro2019}.

The paper is organized as follows: first we review how (weak) disorder 
can be included within
an Euclidean functional formulation for a system of  
interacting bosons in a generic dimension $d$.
Then we move to compute the thermodynamic potential up to the 
Gaussian level in quantum and thermal
fluctuations, by taking into account both the presence of an 
uncorrelated quenched disorder
and a finite-range interaction between the atoms. 
In order to simplify the computational burden, we focus on point-like 
defects. The case $d=3$ is analyzed in more detail
and explicit results are given also for the thermal depletion of 
the superfluid density.

\section{Modelling the disorder}
\label{sec:II}

In this paper, we study the interplay between disorder, 
non-local interaction and fluctuations in a quantum gas made of 
bosonic particles. To fulfill this task, 
our analysis will be carried on within the functional integration framework.

In order to properly construct the path integral for an ensemble 
of identical atomic bosons with mass $m$, we start from
the second-quantized hamiltonian
\begin{equation}
\begin{aligned}
\hat{K} &= \int_{L^d} d^d\mathbf{r}\; 
\hat{\psi}^{\dagger}(\mathbf{r})\bigg[ -\frac{\hbar^2\nabla^2}{2m}
-\mu - U_{D}(\mathbf{r}) \bigg]\hat{\psi} (\mathbf{r}) \\
& + \frac{1}{2} \int_{L^d} d^d\mathbf{r}\,d^d\mathbf{r}'
\hat{\psi}^{\dagger}(\mathbf{r})\hat{\psi}^{\dagger}(\mathbf{r}') 
V(|\mathbf{r}-\mathbf{r'}|)\hat{\psi}(\mathbf{r}')
\hat{\psi}(\mathbf{r})
\end{aligned}
\label{hamiltoniana seconda quantizzazione}
\end{equation}
where  $V(r)$ is the central two-atom interaction potential 
(with $r = |\mathbf{r}|$), while $U_D(\mathbf{r})$ is the random 
external potential,  which takes into account the effect of disordered 
environment. The field operators in 
Eq. \eqref{hamiltoniana seconda quantizzazione} 
respect the usual bosonic commutation relation
$[\hat{\psi}(\mathbf{r}),\hat{\psi}^{\dagger}(\mathbf{r}')]
=\delta^{(d)}(\mathbf{r}-\mathbf{r}')$.
The spatial integration is taken over a large $d-$dimensional 
 ipercubic volume $L^d$; at the end of 
calculation, if interested in the thermodynamic limit, 
we have to take the limit $L\rightarrow \infty$. 

Given Eq. \eqref{hamiltoniana seconda quantizzazione} 
and adopting the functional intergation formalism, 
we can represent the
partition function of the system as \cite{altland-book}
\begin{equation}
\mathcal{Z} = e^{-\beta\Omega} =  \int\mathcal{D}[\psi,\psi^*] \exp\bigg\lbrace
-\frac{1}{\hbar}S_E[\psi,\psi^*]
\bigg\rbrace
\label{partition function def}
\end{equation}
where $\Omega$ is the thermodynamic potential, 
$\psi(\mathbf{r},\tau)$ is the coarse-grained complex field 
modelling the bosons, $\mathcal{D}[\psi,\psi^*]$ is the measure 
of the functional integration over $\psi(\mathbf{r},\tau)$ and 
$\psi^*(\mathbf{r},\tau)$, and
$S_E[\psi,\psi^*]$ denotes the Euclidean (i.e. imaginary time) action defined as
\begin{equation}
S_E[\psi,\psi^*] = \int_0^{\beta\hbar}d\tau \int_{L^d} d^d{\mathbf{r}} \big[
\psi^*\,\hbar\partial_{\tau}\psi + K(\psi,\psi^*)
\big]\;.
\label{euclidean action def}
\end{equation} 
In the equation above $\beta=1/(k_B T)$, $k_B$ being the Boltzmann 
constant and $T$ the absolute temperature. 

It is worth to characterize the external random 
potential $U_D({\bf r})$ from a statistical point of view. 
This implies that its features can be extracted from a probability 
density function (PDF) according to which single realizations 
of a disordered environment are distributed. 
A reasonable choice for $U_D(\mathbf{r})$ can be a centered 
(i.e. zero average) Gaussian as the following one
\begin{equation}
\mathcal{P}[U_D] \propto \exp\bigg\lbrace -\frac{1}{2}
\int d^d\mathbf{r}d^d\mathbf{r}'
U_D(\mathbf{r}) 
\Delta^{-1}(\mathbf{r},\mathbf{r}')
U_D(\mathbf{r}')
\bigg\rbrace\;
\label{gaussian pdf for disorder}
\end{equation}
with a proper normalization factor which is, however, not
important in the proceeding \cite{altland-book}. Indeed, we will mostly 
need the correlator
\begin{equation}
\braket{U_D(\mathbf{r})U_D(\mathbf{r}')}_{\text{dis}} = 
\Delta(\mathbf{r},\mathbf{r'})
\label{correlatore def}
\end{equation}
where $\braket{\ldots}_{\text{dis}}$ has to be intended as
\begin{equation}
\braket{\ldots}_{\text{dis}} = \int \mathcal{D}[U_D]\,\mathcal{P}[U_D]\,(\ldots)\;.
\label{disorder average def}
\end{equation}
From Eq. \eqref{gaussian pdf for disorder} it is clear that we are 
assuming that the characteristic time scale of $U_D(\mathbf{r})$ 
is infinitely long compared to the other ones; we are then restricting 
ourselves to the (important) case of a \textit{quenched} disordered 
environment. Obviously, this assumption can be relaxed 
and the random potential can be taken as time- or temperature-dependent
\cite{tauber1997}. 

Among Gaussian-distributed random configurations, the most simple situation
is provided by potentials with null correlation length. They all have a
$\delta$-like correlator, such
\begin{equation}
{ \Delta}(\mathbf{r},\mathbf{r}') = 
\Delta(\mathbf{r})\; \delta^{(d)}(\mathbf{r}-\mathbf{r}')\;.
\label{correlator zero correlation length}
\end{equation}
Despite being an extremely simplifying assumption, the equation above 
suits well with disorder generated by static and point-like 
bosonic impurities \cite{tauber1997}. More complicated 
 disordered configurations are  discussed, 
for instance, in \cite{falco2009,tauber1997}.

\section{Disorder and finite-range interactions}

\subsection{The finite-range effective potential}

The crucial issue now consists in a proper choice of 
the two-body interaction. Since the first seminal experiments 
with the alkali atoms in the degenerate regime, it appeared that 
reliable analysis and predictions can be carried on by considering 
a contact interactions (i.e. zero-range) $V\big(r\big) = g_0\, \delta^{(3)}(r)$,
where $r=|\mathbf{r}|$ and $g_0=\int d^d{\bf r} V(r)$ is the 
interaction strength. 
This approximation leads to a universal thermodynamics, 
where all the relevant equilibrium features of the system depend 
only on the coupling parameter \cite{haugset-1997,andersen-2004}. 

However, thanks to Feshbach resonances, 
it is possible to explore regimes where deviations
from universalities become relevant
\cite{braaten-2001,gao-2003,sgarlata-2015,cappellaro-2017,
tononi-2018}.
Moreover, in \cite{braaten-1999} it was already pointed out that
some divergences in the perturbative expansion of the thermodynamic 
potential of Bose gases can be healed by involving
additional features of the interaction potential such as its 
characteristic range.

In order to analyze the first correction to the thermodynamic 
properties of the Bose gas due to a finite-range interaction, 
we begin by considering the following low-momentum expansion 
\begin{equation}
\tilde{V}(q) = g_0 + g_2 q^2 + \mathcal{O}(q^4)\;,
\label{finite range interaction}
\end{equation}
of the Fourier transform ${\tilde V}(q)$ 
of two-body interaction potential $V(r)$, 
where $g_0=\tilde{V}(0)$ and $g_0=(1/2)\tilde{V}''(0)$.

By making use of the effective field theory techniques \cite{braaten-1999} 
and of the T-matrix formulation
of two-atoms collisions \cite{stoof-book}, it is possible to
define the coupling constants $g_0$ and $g_2$ in the equation above in terms of 
measurable scattering parameters. As discussed in detail 
in Ref. \cite{cappellaro-2017}, one obtains 
\begin{equation}
g_0 = \frac{4\pi\hbar^2}{m}a_s \;, \quad g_2 = \frac{2\pi \hbar^2}{m} 
a_s^2\,r_s\;.
\label{coupling constants}
\end{equation}
It is worth recalling that this recipe can be likewise applied in 
lower dimensions, namely $d=1$ and 
(with some technical complications) in $d=2$ \cite{tononi-2018}. 

As mentioned above, thanks to Feshbach resonances, one can explore a wide range of
values for the $s$-wave scattering length by tuning an external magnetic fields.
The effective range expansion responsible
for Eqs. \eqref{finite range interaction} and \eqref{coupling constants}
has to be treated carefully close to narrow resonances or zero crossings
of $a_s$ as a function of the magnetic field $B$.
Rigorously speaking, close to a narrow resonance, a multichannel effective theory
should be required, as detailed in \cite{bruun-2005}, naturally reading to finite-range
effects. Other multi-channel approaches rely upon the formation of formation of
molecular bound states in the regime of resonant interaction 
\cite{kokkelmans-2002,stoof-2005,pricoupenko-2013}. 

However, it has been shown that it is possible to reduce the multi-channel scattering
problem to a single-channel one \cite{gao-2011}, at least for a broad resonance.
The analysis reported in \cite{gao-2011} is further extended
for zero-point crossings and narrow resonances in \cite{blackley-2014}
where it is also compared with multichannel numerical outcomes.
Remarkably, for the isolated resonances in $^{39}$K and $^{133}$Cs 
a very good agreement is found, providing
a very good single-channel approximation.  

\subsection{$\delta$-correlated Gaussian disorder: 
perturbative analysis}

At the mean-field level, 
the space-time translationally-invariant 
ground state of the system can be described by a 
constant $v$, namely $\psi(\mathbf{r},\tau) = v$. 
For the sake of simplicity the constant $v$ is 
taken real. By inserting it in Eq. \eqref{euclidean action def}, 
the mean-field thermodynamic potential reads
\begin{equation}
\frac{\Omega_{\text{mf}}(\mu,v)}{L^d} = -\big[\mu 
+\braket{U_D}_{\text{dis}}\big]  v^2 + 
\frac{1}{2}\tilde{V}(0)\; v^4\;,
\label{mean field granpot}
\end{equation} 
where we have already considered the disorder average defined 
in Eq. \eqref{disorder average def}. However, in the following, we are going to consider 
 the case of $\delta$-correlated centered Gaussian 
disorder, so $\braket{U_D}_{\text{dis}} = 0$.
Eq. \eqref{mean field granpot} is stationarized only by $v=0$ for $\mu < 0$, 
but $v$ acquires a non-zero value for $\mu >0$:
\begin{equation}
v^2 =  \frac{\mu}{\tilde{V}(0)}\; .  
\label{non zero order parameter}
\end{equation}
This fact obviously signals the occurring of the superfluid transition, 
where a $U(1)$ symmetry is spontaneously broken. 

It is important to remark 
that a $\delta$-correlated does not affect the mean-field
picture of the homogeneous and stationary ground state. 
Moreover, if we consider the interaction potential as given 
by Eq. \eqref{finite range interaction}, we immediately realize 
that only $g_0={\tilde V}(0)$ appears up to this level 
of approximation. As a consequence, in order to explore 
the role of disorder and nonlocal interactions, one has to 
consider fluctuations, both quantum and thermal, above the ground state 
given by Eq. \eqref{non zero order parameter}.

Let us then consider the following shift of the field 
\begin{equation}
\psi(\mathbf{r},\tau) = v + \eta(\mathbf{r},\tau)
\label{shift of the field}
\end{equation}
with $\eta(\mathbf{r},\tau)$ being the complex fluctuating field.

By replacing Eq. \eqref{shift of the field} in the 
Euclidean action $S_E[\psi,\psi^*]$ and retaining terms up to the 
quadratic (Gaussian) level in $\eta$ and $\eta^*$, the partition function 
defined in Eq. \eqref{partition function def} can be factorized in 
\begin{equation}
\begin{aligned}
\mathcal{Z}[U_D] & \simeq e^{-\beta\Omega_{\text{mf}}} 
\int \mathcal{D}[\eta,\eta^*]
e^{-\frac{1}{\hbar} \big( S_g^{\text{(pure)}}[v,\eta]   + S_g^{\text{(dis)}}[v,\eta,U_D]\big)}
\end{aligned}
\label{partition function fattorizzata}
\end{equation}
where $\Omega_{\text{mf}}$ is given by Eq. \eqref{mean field granpot},
$S_g^{\text{(pure)}}$ is the Euclidean action describing the periodic 
imaginary-time trajectories of the fluctuating fields $\eta$ and $\eta^*$, 
while the disorder contribution is encoded in $S_g^{\text{(dis)}}$. 

Within our perturbative scheme, $S_g^{\text{(pure)}}$ reads the same 
expression for a system not subject to an external disordered environment. 
In the Fourier space, this implies that
\begin{equation}
S_g^{\text{(pure)}} = \frac{\hbar}{2} \sum_{\mathbf{q},\omega_n} 
\Psi^{\dagger}(\mathbf{q},\omega_n)
\mathbb{M}(\mathbf{q},\omega_n,v) \Psi(\mathbf{q},\omega_n)\;.
\label{azione gaussiana pura}
\end{equation}
with the spinor $\Psi^{\dagger}(\mathbf{q},\omega_n) = [
{{\tilde \eta}}^*(\mathbf{q},\omega_n),
{{\tilde \eta}}(-\mathbf{q},-\omega_n)]$
 and 
\begin{widetext}
\begin{equation}
\mathbb{M}(\mathbf{q},\omega_n,v) = \beta
\begin{pmatrix}
-i\hbar\omega_n + \dfrac{\hbar^2 q^2}{2m} -\mu +g_0v^2 +v^2(g_0+g_2q^2)& 
v^2(g_0+g_2q^2) \\
 &  \\
v^2(g_0+g_2q^2) &  i\hbar\omega_n + \dfrac{\hbar^2 q^2}{2m} -\mu +g_0v^2 
+v^2(g_0+g_2q^2)\\ 
\end{pmatrix}
\label{inverse of gaussian propagator}
\end{equation}
\end{widetext}
is the inverse of Gaussian propagator. In this framework, the discrete 
frequencies labelled by $n \in \mathbb{Z}$ are the usual Matsubara ones, 
defined as $\omega_n= 2\pi n /(\beta\hbar)$ \cite{altland-book}.

On the other hand, the disorder contribution $S_g^{\text{(dis)}}$ can be 
written down as \cite{schakel1997}
\begin{equation}
S_g^{\text{(dis)}} = -v\int_0^{\beta\hbar}d\tau\int d^d\mathbf{r}
\,U_D(\mathbf{r}) \big(1,1) \cdot
\begin{pmatrix}
\eta(\mathbf{r},t) \\
\eta^*(\mathbf{r},t)
\end{pmatrix}\;.
\label{disorder contribution gaussian action}
\end{equation}
Since Eq. \eqref{partition function fattorizzata} has, by construction, 
a Gaussian structure, the functional integration over $\eta$ and $\eta^*$ 
can be performed exactly, reading, in Fourier space \cite{cappellaro2019},
\begin{equation}
\begin{aligned}
\mathcal{Z}[U_D] & \simeq e^{-\beta\Omega_{\text{mf}}}
\exp\bigg\lbrace
- \frac{1}{2}\sum_{\mathbf{q},\omega_n}\log\big[ \det\mathbb{M}(q,\omega_n)\big]
\bigg\rbrace  \\
& \times \exp\bigg\lbrace 
\frac{1}{2}\sum_{\mathbf{q}} {|{\tilde U}_D(\mathbf{q})|^2}
(v,v)\mathbb{M}^{-1}(\mathbf{q},0)
\begin{pmatrix}
v \\ 
v
\end{pmatrix}
\bigg\rbrace\; , 
\end{aligned}
\label{partizione eta integrato}
\end{equation}
where ${\tilde U}_D({\bf q})$ is the Fourier transform of the 
random potential $U_D({\bf r})$. 
Let us remark that, in the equation above, the propagator $\mathbb{M}$ 
is computed at $\omega_{n}=0$, so there is no dependency on the temperature 
in the disorder contribution. As stated in Sec. \ref{sec:II}, 
this is due to the assumption of a quenched disorder, whose 
characteristic features are frozen compared to other (quantum and thermally) 
fluctuating quantities.

The pure contribution of Gaussian fluctuations leads us to 
\begin{equation}
\Omega_g^{\text{(pure)}}(\mu,v) = \frac{1}{2\beta} \sum_{\mathbf{q},\omega_n}\log\big[
\beta^2\big(\hbar^2\omega_n^2 +E_q^2\big)
\big]
\label{gaussiano puro con frequenze}
\end{equation}
with
\begin{equation}
\begin{aligned}
E_q^2 = \bigg[\frac{\hbar^2q^2}{2m} -\mu +g_0v^2 + v^2\tilde{V}(q)
\bigg]^2
-v^4\tilde{V}^2(q)
\end{aligned}
\label{excitation spectrum}
\end{equation}
being the elementary excitation spectrum over the uniform ground state.
By performing the Matsubara summation in Eq. 
\eqref{gaussiano puro con frequenze}
we can identify the quantum and thermal contribution to $\Omega_g^{\text{(pure)}}$:
\begin{equation}
\Omega_g^{\text{(pure)}} = \frac{1}{2}\sum_{\mathbf{q}} E_q +
\frac{1}{\beta} \sum_{\mathbf{q}} \log\big( 1- e^{-\beta E_q}\big)\;.
\label{granpot puro def}
\end{equation}
Clearly, $\Omega_g^{\text{(pure)}}$ is independent from $U_D(\mathbf{r})$, so 
Eq. \eqref{granpot puro def} is left untouched by the (functional) disorder 
average introduced in Eq. \eqref{disorder average def}, i.e.
$\braket{\Omega_g^{\text{(pure)}}}_{\text{dis}}=\Omega_g^{\text{(pure)}}$. 
Obviously, this is not true for disorder term. 
Indeed, if we take the logarithm of 
the second line of Eq. \eqref{partition function fattorizzata} we get the
corresponding
\begin{equation}
\Omega_g^{\text{(dis)}} = - \frac{v^2}{2}\sum_{\mathbf{q}}
{ |{\tilde U}_D(\mathbf{q})|^2} (1,1)\cdot\mathbb{M}^{-1}(\mathbf{q},0) 
\begin{pmatrix}
1 \\
1
\end{pmatrix} \;.
\label{granpot gaussiano disordine}
\end{equation}
Under the assumption of a Gaussian distribution disorder, 
$\braket{\Omega_g^{\text{(dis)}}}_{\text{dis}}$ can be easily computed by 
 using Eq. \eqref{correlatore def}, which gives
\begin{equation}
\int \mathcal{D}[U_D] \mathcal{P}[U_D] |U_D(\mathbf{q})|^2 =
\braket{{ |{\tilde U}_D(\mathbf{q})|^2}}_{\text{dis}} 
=  {\tilde \Delta}(\mathbf{q}) \; , 
\label{correlatore dis fourier}
\end{equation}
 where ${\tilde \Delta}({\bf q})$ is the Fourier transform of 
the delta-correlated disorder field $\Delta({\bf r})$.
In this way we obtain  
\begin{equation}
\braket{\Omega_g^{\text{(dis)}}}_{\text{dis}} = - \frac{v^2}{2}\sum_{\mathbf{q}}
{ {\tilde \Delta}(\mathbf{q})}\; 
(1,1) \cdot \mathbb{M}^{-1}(\mathbf{q},0) 
\begin{pmatrix}
1 \\
1 
\end{pmatrix}\;.
\label{granpot dis mediato}
\end{equation}
The inverse of Eq. \eqref{inverse of gaussian propagator} is immediate 
to compute, such that, at the continuum limit 
$\sum_{\mathbf{q}} \rightarrow L^d \int d^d\mathbf{q}/(2\pi)^d$,
\begin{equation}
\begin{aligned}
\frac{\braket{\Omega_g^{\text{(dis)}}}_{\text{dis}}}{L^d} & = 
- v^2 \int \frac{d^d\mathbf{q}}{(2\pi)^d}
\dfrac{{ {\tilde \Delta}(\mathbf{q})}}{\dfrac{\hbar^2q^2}{2m} 
+ 2v^2g_2\,q^2-\mu + 3g_0v^2}\;.
\end{aligned}
\label{granpot dis al continuo}
\end{equation}
In the following section we derive analytical  
and numerical results for
the disorder contribution to the thermodynamic potential 
provided Eq. \eqref{granpot dis al continuo} as a starting point.

\section{Point-like defects}

\subsection{Disorder contributions in d dimensions}

In order to proceed with our calculation, it is necessary to specify the 
spatial behaviour (or, equivalently, the dependence on $\mathbf{q}$) 
of the disorder correlator $\Delta(\mathbf{\mathbf{r}})$. 
For point-like defects, Eq. \eqref{correlator zero correlation length} 
is further simplified assuming 
\begin{equation}
{{\tilde \Delta}(\mathbf{q}) = \Delta } \; , 
\label{pointlike defects}
\end{equation}
where $\Delta$ is a real constant. 
As mentioned earlier, despite this simple assumption, the system, 
in its superfluid phase is nevertheless crucially affected by the presence 
of a disordered external environment.

In order to highlight this feature,  given Eq. \eqref{pointlike defects},
let us recast Eq. \eqref{granpot dis al continuo} as
\begin{equation}
\frac{\braket{\Omega_g^{\text{(dis)}}}_{\text{dis}}}{L^d} 
= - \frac{\Delta \,v^2}{\lambda(v^2,g_2)}
\int\frac{d^d\mathbf{q}}{(2\pi)^d}\; \frac{1}{\dfrac{\hbar^2q^2}{2m} 
+ \dfrac{3g_0 v^2-\mu} {\lambda(v^2,g_2)}}
\label{granpot riscritto}
\end{equation}
where, in analogy with the notation used in \cite{cappellaro-2017}, 
we have defined
\begin{equation}
\lambda(v^2,g_2) = 1 + \frac{4mg_2}{\hbar^2}v^2 \;.
\label{lambda def}
\end{equation}
At this point, it is immediate to realize that the disorder contribution
$\braket{\Omega_g^{\text{(dis)}}}_{\text{dis}}/L^d$ diverges for $d\geq 2$.
However, it has been extensively shown (see 
\cite{andersen-2004,braaten-2001,zeidler-book} for technical details)
that (finite) meaningful information may be extracted by means of the 
so-called dimensional regularization. 

The key point of this method consists in performing the integration 
in Eq. \eqref{granpot riscritto} in a generic complex dimension 
$d_{\epsilon} = d-\epsilon$ with values of $\epsilon \in \mathbb{C}$ 
for which the result is convergent. 
The last step requires an analytical continuation back to the physical 
dimension we are interested in, i.e. one to consider a proper limit 
procedure such that $\epsilon \rightarrow 0$. 
Within this regularization framework, the following result holds
\begin{equation}
\mathcal{I}(d) = \int \frac{d^d\mathbf{q}}{(2\pi)^d} \, \frac{1}{q^2 + M^2}
= \frac{\Gamma(1-d/2)}{2^d\pi^{d/2}} M^{d-2} \;,
\label{regolarizzazione dimensionale risultato}
\end{equation}
$\Gamma(z)$ being the Euler's Gamma function.
As a consequence, Eq. \eqref{granpot riscritto} reads
\begin{equation}
\begin{aligned}
\frac{\braket{\Omega_g^{\text{(dis)}}}_{\text{dis}}}{L^d} 
& = - \frac{\Gamma(1-d/2)}{(2\pi)^{d/2}}
\bigg(\frac{m}{\hbar^2}\bigg)^{d/2}\\
& \qquad \qquad \frac{\Delta\,v^2 \big(3g_0\,v^2 -\mu \big)^{d/2-1}}
{\lambda^{d/2}(v^2,g_2)}\;.
\end{aligned}
\label{granpot integrale fatto d generico}
\end{equation}

Moving from Eq. \eqref{granpot integrale fatto d generico}, 
a simple derivative leads us to the disorder contribution to the total 
density as a function of the condensed one. More technically,
\begin{equation}
\begin{aligned}
n_g^{\text{(dis)}}(n_0)  & = - \frac{1}{L^d} \frac{\partial}{\partial \mu}
\braket{\Omega^{(0)}_{\text{dis}}(v,\mu)}_{\text{dis}} 
\bigg|_{\substack{\mu = g_0 n_0 \\ v^2 = n_0}} \\
& \\
& = \frac{\Gamma(2-d/2)}{4\pi^{d/2}}\bigg(\frac{m}{\hbar^2}\bigg)^{d/2}
\frac{\Delta\, g_{0}^{d/2-2}\, n_0^{d/2-1}}{\lambda^{d/2}(n_0,g_2)} \;.
\end{aligned}
\label{calcolo densita disordine}
\end{equation}
It is fundamental to underline that, following a perturbative scheme, 
we have identified $v^2$ as given in Eq. \eqref{non zero order parameter} 
with the density $n_0$ of condensed atoms. Thus, our approach provides 
an implicit expression for the condensate density of the system.
Since the pure and the disorder contributions to the thermodynamic potential 
are additive, as evident from Eqs. \eqref{partition function fattorizzata} 
and \eqref{granpot puro def}, the same occurs for the contributions 
to the total number density, i.e.
\begin{equation}
n = n_0 + n_g^{(0)}(n) + n_g^{(T)}(n) + n_g^{\text{(dis)}}(n) \;.
\label{total density}
\end{equation}
Let us also point out that, within a perturbative approach for the weakly 
interacting system,
it is possible to approximate $n_0 \approx$ in the fluctuation 
corrections $n_g^{(0)}(n)$,
$n_g^{(T)}(n)$ and $n_g^{\text{(dis)}}(n)$. 
Thus, in the following, we are going to approximate $n_0$ with $n$ in 
the perturbatively-computed fluctuation contributions.
In Eq. \eqref{total density}, $n_g^{(0)}$ and $n_g^{(T)}$ are 
computed similarly to 
Eq. \eqref{calcolo densita disordine}: first, one has to extract the 
regularized Gaussian corrections from Eq. Eq. \eqref{granpot puro def}, 
then derive with respect to the chemical potential.

\begin{figure*}[ht!]
\subfigure{
\includegraphics[scale=0.35]{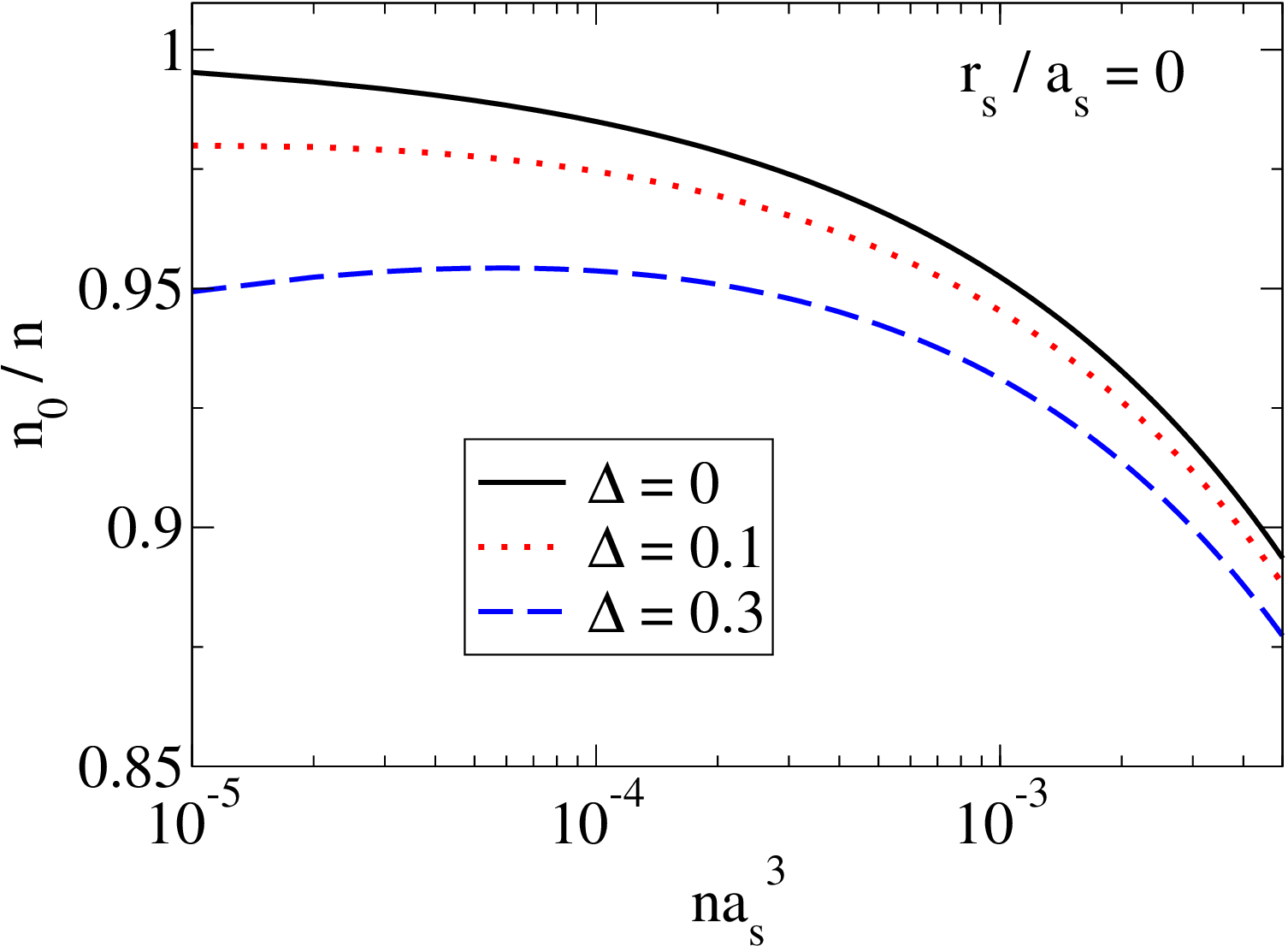}
\includegraphics[scale=0.35]{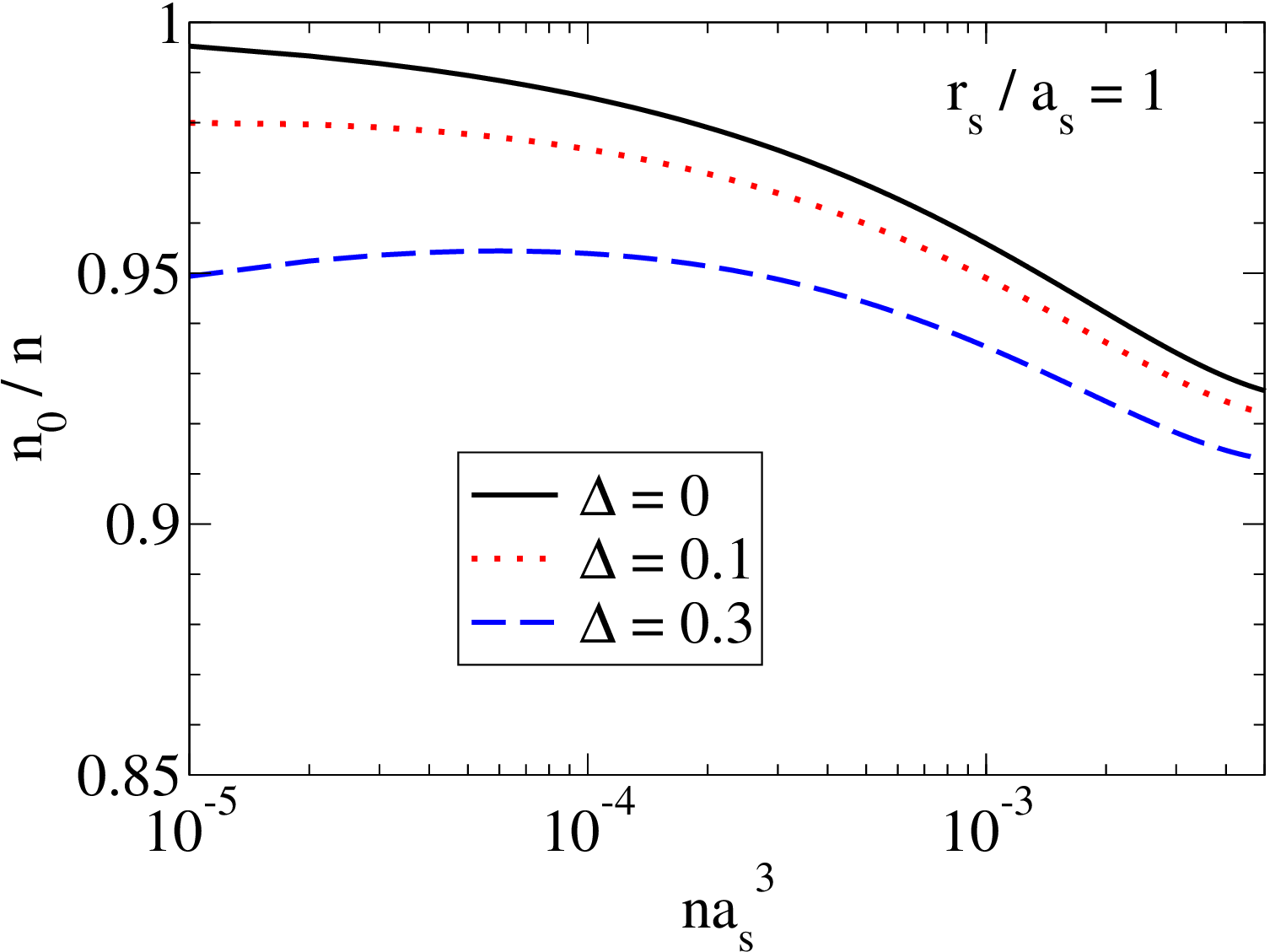}
}
\par\bigskip
\subfigure{
\includegraphics[scale=0.35]{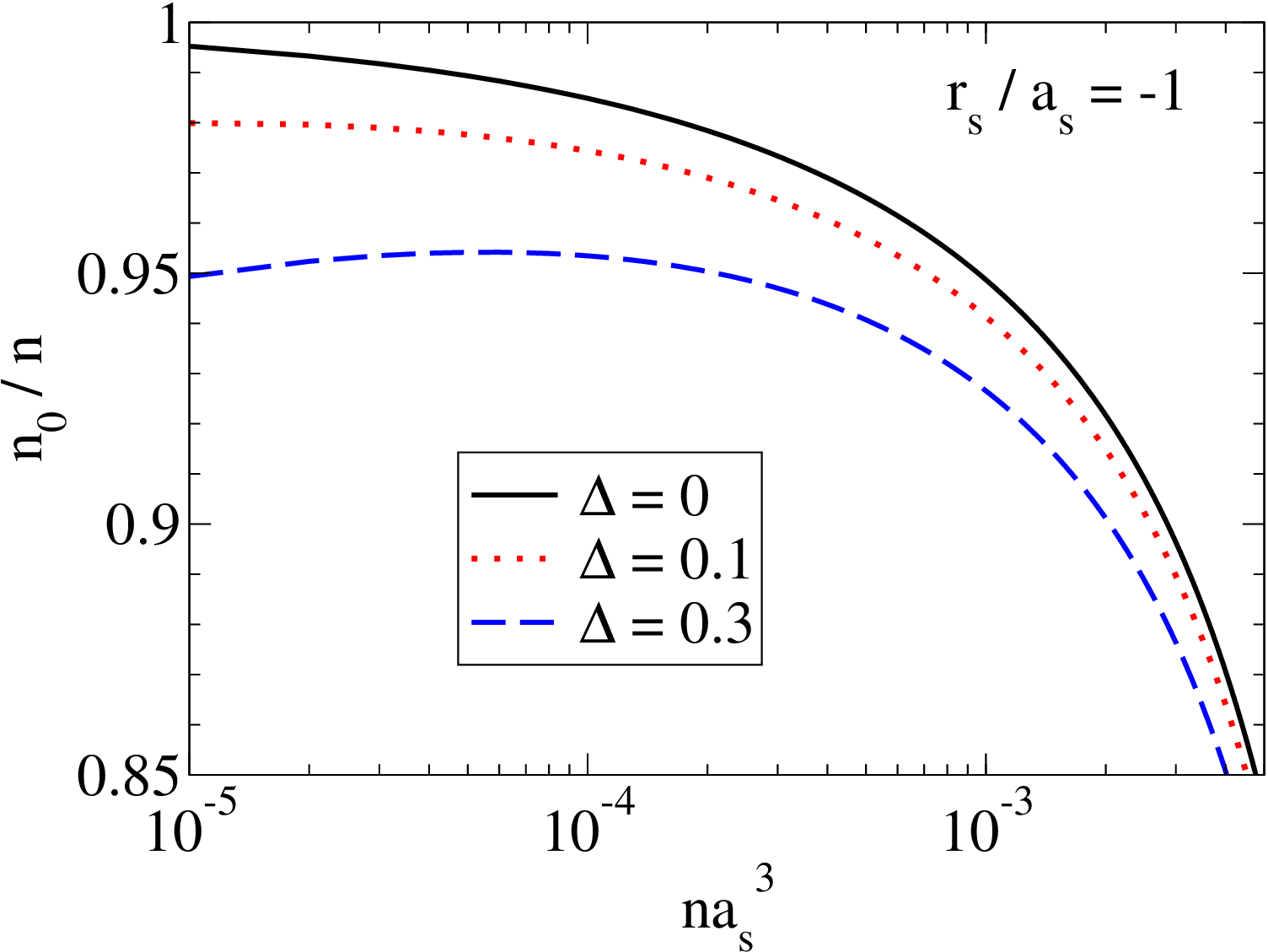}
}
\caption{
Condensate fraction $n_0/n$ at zero temperature as a function 
of the dimensionless gas parameter $na_s^3$ for increasing disorder strength $\Delta$. 
The three panels correspond to different values of the
 ratio $r_s/a_s$ between the effective range $r_s$ and the 
scattering length $a_s$. 
In the legend, $\Delta$ is expressed in units of 
$E_B^2/n$ with $E_B = \hbar^2n^{2/3}/m$.
The curves are obtained from  Eqs. 
\eqref{density pure system} and \eqref{frazcondo},
which provide a reliable thermodynamic picture 
of the system within the range of diluteness 
(i.e. the values of $na_s^3$) chosen for the horizontal axis.
}
\label{fig1}
\end{figure*}

In \cite{tononi-2018}, an extensive analysis addressed this topic 
for the pure system. Being interested in the role played by the disorder, 
we report here only the final result, i.e.
\begin{widetext}
\begin{equation}
\begin{aligned}
n_g^{(0)}(n) & = \frac{1}{4\pi^{d/2}\Gamma(d/2)} 
\bigg(\frac{m}{\hbar^2}\bigg)^{d/2}
\frac{(g_0 n)^{d/2}}{\lambda^{(d-1)/2}(n,g_2)}\bigg[
2\,\frac{1+\frac{2m}{\hbar^2}g_2n}{\lambda(n,g_2)}\mathcal{B}
\bigg(\frac{d+1}{2},-\frac{d}{2}\bigg)
+ \mathcal{B}\bigg(\frac{d-1}{2},\frac{2-d}{2} \bigg) 
\bigg] \\
& \\
n_g^{(T)}(n) & = \int \frac{d^d\mathbf{q}}{(2\pi)^d} 
\frac{\dfrac{\hbar^2q^2}{2m}\bigg(1+
	\dfrac{2m}{\hbar^2}g_2n \bigg)+g_0n}
{\sqrt{\dfrac{\hbar^2q^2}{2m}\bigg[ 
		\lambda(n,g_2)\dfrac{\hbar^2q^2}{2m} + 2g_0n
		\bigg]}} \bigg( 
	\frac{1}{e^{\beta E_q(n)}-1}
	\bigg)
\end{aligned}
\label{density pure system}
\end{equation}
\end{widetext}
with $\mathcal{B}(x,y)$ the Euler beta function, which 
can be rewritten, after analytic continuation, 
in terms of the Euler gamma function $\Gamma(x)$ as follows 
$\mathcal{B}(x,y) = \Gamma(x)\Gamma(y)/{\Gamma(x+y)}$. 
$E_q(n)$ is the spectrum of collective excitations given by 
Eq. \eqref{excitation spectrum} with $n_0 \approx n$.

\section{Analytical and numerical results in three dimensions}

We now consider the common situation of bosons moving in three spatial 
dimensions, i.e. $d=3$. In this case, during the calculation of 
$\Omega(v,\mu)$, only power divergences arise. 
Within the dimensional regularization scheme, they are set to zero and 
no explicit renormalization is required \cite{andersen-2004}; 
this holds for both the pure and disorder terms.

\subsection{Condensate density}

By taking the limit $d\rightarrow 3$ in Eqs. 
\eqref{calcolo densita disordine} and \eqref{density pure system}, 
at zero temperature one gets
\begin{eqnarray}
{n_0\over n} &=& 1 - {32\over 3\sqrt{\pi}} \ {\sqrt{n a_s^3}\over 
1+8\pi (n a_s^3) \big(\frac{r_s}{a_s}\big)} \left[ 
{ 1+4\pi (n a_s^3) \big(\frac{r_s}{a_s}\big) \over 
1+8\pi (n a_s^3) \big(\frac{r_s}{a_s}\big) } - {3\over 4} \right] 
\nonumber 
\\
&-& {1\over (4\pi)^{3/2}} \
{\Delta \ (\frac{m^2}{\hbar^4n^{1/3}}) \over \big[1+8\pi (n a_s^3) 
\big(\frac{r_s}{a_s}\big)\big]^{3/2} (n a_s^3)^{1/6}} \; . 
\label{frazcondo}
\end{eqnarray}

In Fig. \ref{fig1} we report the behaviour of the condensate 
fraction $n_0/n$ at $T = 0$ as a function of the gas parameter $na_s^3$ for 
different values of the effective range and increasing the disorder 
correlator $\Delta$. 
We immediately realize that the weight of the finite-range corrections 
is enhanced at higher densities (for a given s-wave scattering length). 
Indeed, from the middle and the right panel of Fig. \ref{fig1} 
one can observe that, depending on the sign of 
the ratio $r_s/a_s$, the condensate is depleted slower (for $r_s/a_s >0$) 
or faster (when the ratio is negative). Moreover, the disorder contribution, 
depending on $r_s/a_s$, is relevant also at very low values of $na_s^3$, 
where the particles seem to be much more sensitive to the presence 
of a disordered external potential. As clearly shown by the dotted red curve 
and the dashed-dotted blue curve in the panels of Fig. \ref{fig1}, 
the joint presence of disorder and finite-range corrections modifies the
condensate depletion of the system.

Despite the corrections reported in Fig. \ref{fig1} are tiny in
magnitude, calculations in the pure case \cite{braaten-2001,gao-2003,cappellaro-2017}
show that the finite-range (Gaussian) contribution to $\Omega(\mu,v)$ removes
an artificial thermodynamic instability and consequently expand the applicability range
of a Gaussian theory. More precisely, by using Eq. \eqref{non zero order parameter}
with $\tilde{V}(0) \rightarrow g_0$ (i.e. $r_s \rightarrow 0$), one
can express the thermodynamic potential as a function of the chemical potential $\mu$.
By recalling that $P(\mu) = - \Omega(\mu)/L^3$ and that the uniform configuration is
stable for $\partial^2_{\mu} P(\mu)>0$, one realizes that the uniform configuration 
is unstable above a critical value of the chemical potential corresponding to 
$(na_s)^3_{c} \simeq 0.004$. In \cite{cappellaro-2017}, analytical results for
a Bose gas made of hard spheres (where $r_s / a_s = 2/3$) are compared to 
corresponding Path-Integral Ground-State MonteCarlo (MC) Simulation \cite{rossi-2013},
finding a very good agreement between theoretical predictions and numerical outcomes.

Moreover, already in \cite{gao-2003}, a modified version of the GPE including corrections
due to Gaussian fluctuations and finite-range interactions was found to reproduce
reasonably well numerical simulations based on Diffusion MC methods, rather than its
zero-range counterpart. Obviously,
refined MC simulation are able to explore the behaviour of (bosonic) quantum
gases well beyond the range of validity of the Gaussian (i.e. Bogoliubov) approach,
at more dense regime and with higher values of the disorder strengths 
\cite{pilati-2010}.

\subsection{Superfluid density}

The interplay between disorder and non-local 
interactions can be effectively understood by analyzing, 
the depletion 
of the superfluid density. According to the Landau phenomenological 
description \cite{landau-book}, in presence
of a superflow, the total density of the system can be split into
\begin{equation}
n = n_s + n_n^{\text{(pure)}}(n) + n_n^{\text{(dis)}}(n)\;,
\label{superfluidity}
\end{equation}
with $n_s$ the superfluid density and $n_n=n_n^{\text{(pure)}}+
n_n^{\text{(dis)}}$ the normal one. 
The depletion of the superfluid is usually driven by the thermal 
activation of the collective excitations. 
\begin{figure*}[ht!]
\centering
\subfigure{
	\includegraphics[scale=0.35]{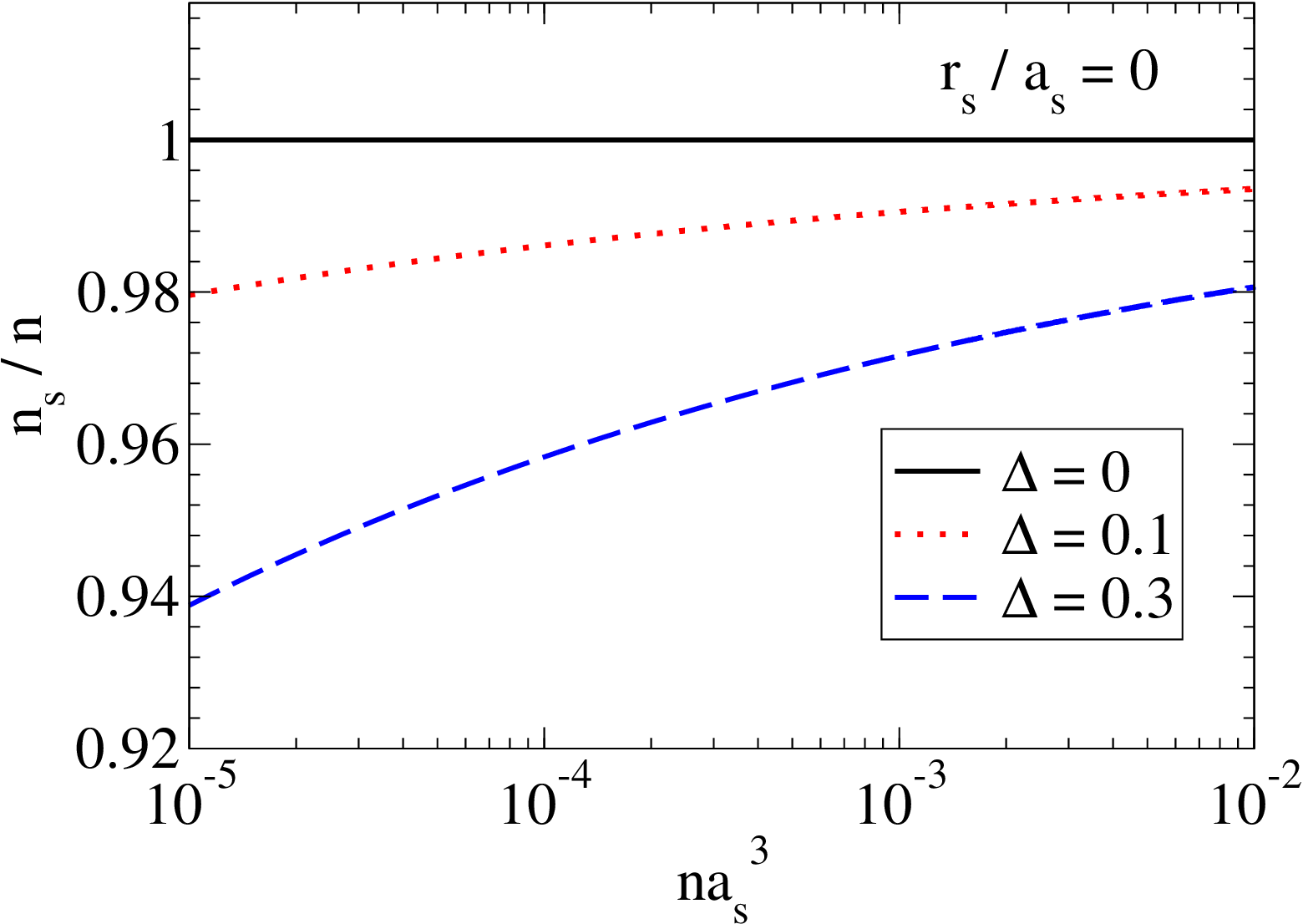}
	\includegraphics[scale=0.35]{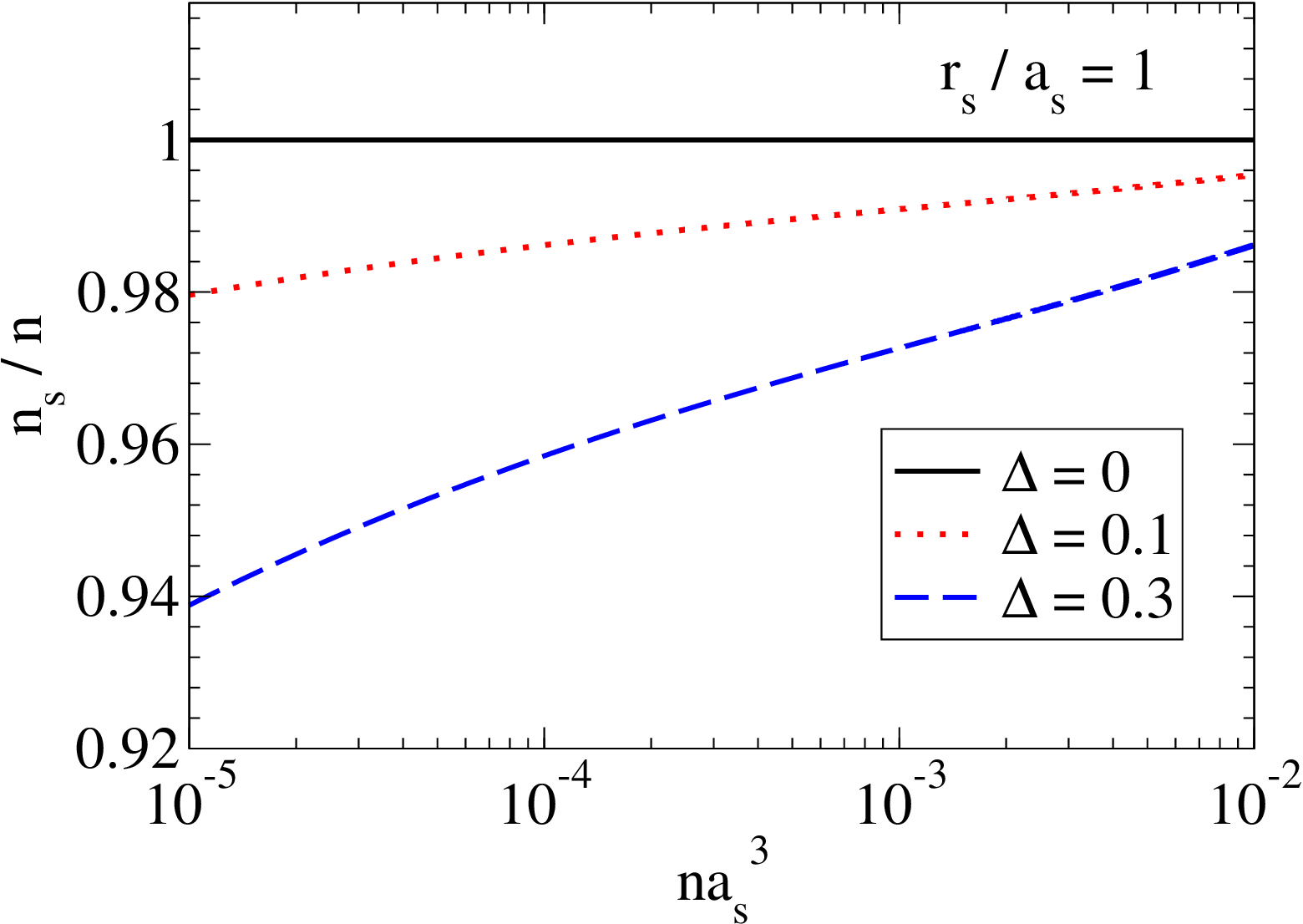}
}
\par\bigskip
\par\bigskip
\subfigure{
	\includegraphics[scale=0.35]{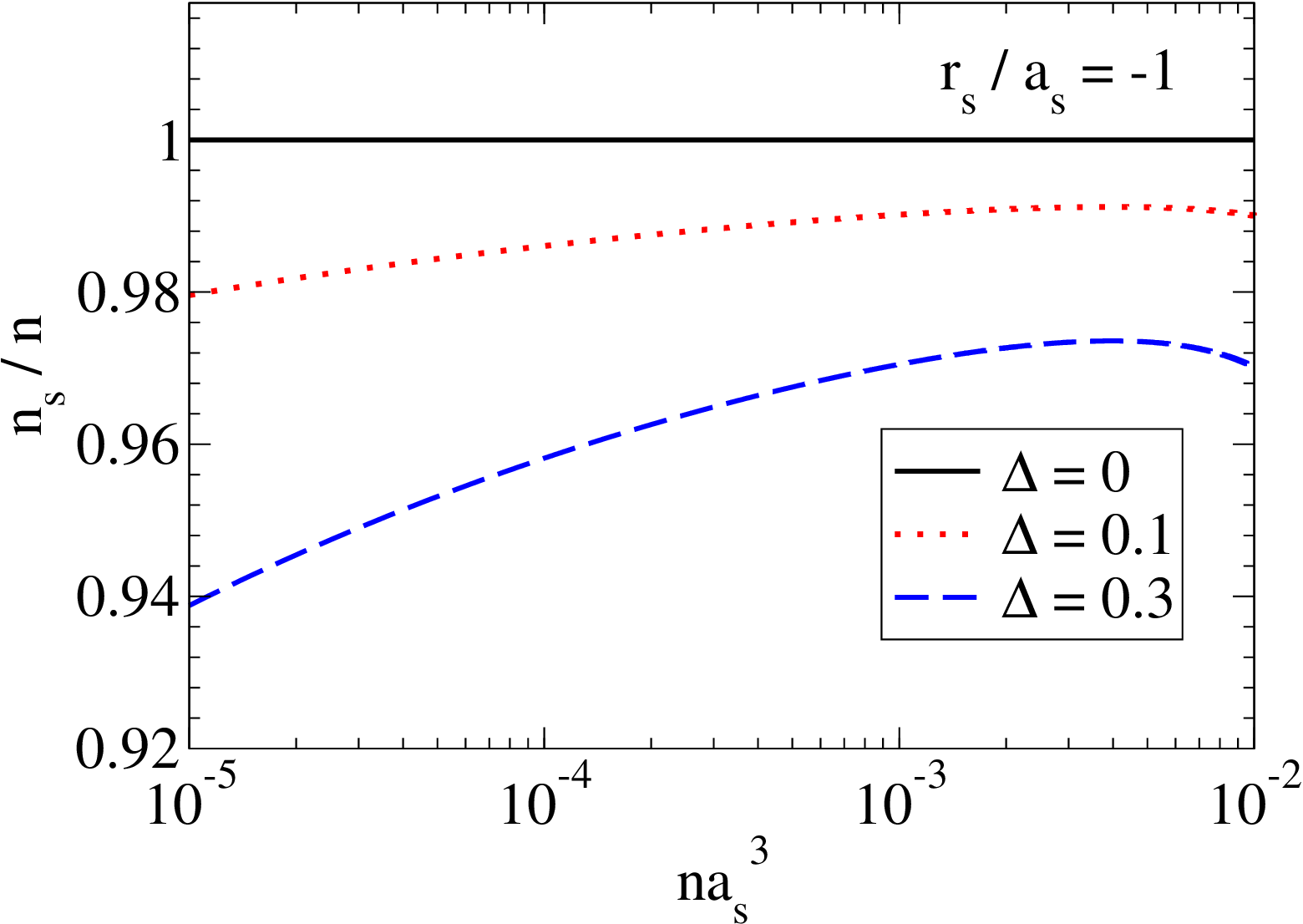}
}
\caption{
Superfluid fraction $n_s/n$ as a function of the dimensionless gas
parameter $na_s^3$, 
for different combinations of $\Delta$ and the ratio $r_s/a_s$. 
As in Fig. \ref{fig1}, the disorder strength 
$\Delta$ is expressed in units of $E_B^2/n$ with $E_B = \hbar^2n^{2/3}/m$ 
and $r_s/a_s$ is the ratio between the effective range $r_s$ and the 
scattering length $a_s$. The curves have been obtained 
from Eq. \eqref{wenzi}.
}
\label{fig2}
\end{figure*}
For instance, in $d=3$ dissipation can occur via density waves 
(i.e. phonons). As detailed in 
\cite{tauber1997,schakel-book} the original intuition of Landau and Khalatnikov
can be adapted to a field-theory analysis, reading the well-known 
formula 
\begin{equation}
n_n^{\text{(pure)}}(n,T) = 
\frac{\beta}{4d}\int \frac{d^d\mathbf{q}}{(2\pi)^d} 
\bigg(\frac{\hbar^2 q^2}{m}\bigg)
\frac{e^{\beta E_q}}{(e^{\beta E_q}-1)^2}
\label{depletion superfluid puro}
\end{equation}
with $E_q$ given, again, by Eq. \eqref{excitation spectrum} with 
$n_0 \approx n$.

However, it is known that the presence of disorder adds another source of depletion, 
also at $T = 0$. For static point-like defects, this disorder 
contribution has a peculiar simple form 
\cite{schakel-book,cappellaro2019}, i.e.
\begin{equation}
n^{\text{(dis)}}_{n}(n) = \frac{4}{d} n_g^{\text{(dis)}}(n)
\label{depletion superfluido disordine}
\end{equation}
where $n_{\text{dis}}(n)$ has been computed in 
Eq. \eqref{calcolo densita disordine}. 
Thus, if $n$ is known, 
Eqs. \eqref{superfluidity}-\eqref{depletion superfluido disordine}
provide an easy way to extract the superfluid fraction of the system. 
In the case $d=3$ one finds 
\beq 
{n_s\over n} = 1 - {1\over 6\pi^{3/2}}\, 
{\Delta \ \bigg(\dfrac{m^2}{\hbar^4n^{1/3}}\bigg) \over \bigg[1+8\pi (n a_s^3) 
\bigg(\dfrac{r_s}{a_s}\bigg)\bigg]^{3/2} (n a_s^3)^{1/6}} \; .
\label{wenzi}
\eeq

In Fig. \ref{fig2} we report the behaviour of $n_s/n$, according to the 
equation above as a function of the gas parameter $na_s^3$. 
In the left panel of Fig. \ref{fig2} we include the results 
in absence of a finite-range interaction, in order to better 
understand the case of $r_s \neq 0$.Indeed, for a positive value of the ratio
$r_s/a_s$ (middle panel), the superfluid fraction does not seem to be 
significantly affected. On the contrary, for a negative value of the 
effective range $r_s$, the left panel of Fig. \ref{fig2} shows that the 
behavior of $n_s/n$ is no more monotonous within the range of values 
we have considered for $na_s^3$. This an example of the interplay between 
between disorder and finite-range interactions, both affecting a 
relevant transport quantity such as the superfluid fraction of the system.

For the sake of completeness, we notice that from Eq. (\ref{wenzi}) 
one can also deduces the critical disorder strength 
\beq 
\Delta_c = 6\pi^{3/2} {\hbar^4n^{1/3}\over m^2} 
{\bigg[1+8\pi (n a_s^3) \bigg({r_s\over a_s}\bigg)\bigg]^{3/2} (n a_s^3)^{1/6}}
\eeq
above which the superfluidity is destroyed despite the absence of 
thermal excitations. The formula shows that the effective range $r_s$ 
induces a nonlinear shift on $\Delta_c$. 

\section{Conclusions}

We have considered a bosonic system with a finite-range two-body interaction 
placed in a disordered environment. 
We have investigated the superfluid phase of this system and, 
according to our perturbative field-theoretical analysis valid in 
any spatial dimension $d$, we have computed explicitly, 
Eqs. (\ref{frazcondo}) and (\ref{wenzi}), 
the modified depletion of the condensate 
and superfluid density in the three-dimensional case. 
Our results show that disorder and 
non-local interactions simultaneously modify the contribution of quantum 
and thermal fluctuations, leading to nonuniversal corrections. 
These theoretical predictions become very important 
when the s-wave effective range $r_s$ of the inter-atomic potential 
is of the same order (or larger) with respect to the s-wave 
scattering length $a_s$. 
This regime can be achieve by approaching a zero-point crossing
of the scattering where the effective range (otherwise constant and $\sim 10 a_0$, with
$a_0$ the Bohr's radius) may vary and change its sign. It has been pointed out
\cite{blackley-2014} that, in this regime,
a single-channel approximation \cite{gao-2011} provides quite reliable results.

In lower spatial dimensions quantum and thermal fluctuations are strongly 
enhanced and a proper characterization of their 
contribution is mandatory, also in terms of 
interaction parameters. Moreover, the resulting thermodynamic picture 
may serve as a starting point to build an effective Gross-Pitaveskii-like 
equation in the spirit of the local-density-approximation. This has been done, 
for instance, with strongly magnetic atoms \cite{wachtler2016,bisset2016},
binary mixtures \cite{malomed2018} or spin-orbit coupling \cite{ming2018}.
However, all these papers does not consider a disordered environment 
which crucially alters the superfluid dynamics 
of a condensed systems. The physical picture becomes more richer if 
we allow the possibility of a localized phase but first we need to
understand the interplay between disorder and non-local interaction in 
the fluctuations contribution, which may help (or not) in driving 
the system towards the superfluid-to-localized transition. 

\end{document}